\documentclass[%
 reprint,
superscriptaddress,
 amsmath,amssymb,
 aps,
prr,
longbibliography,
]{revtex4-1} %4-1

\usepackage{graphicx}% Include figure files
\usepackage{dcolumn}% Align table columns on decimal point
\usepackage{bm}% bold math
\usepackage{xcolor}
\usepackage{mathtools}
\usepackage{upgreek}
\begin{document}

%\preprint{APS/123-QED}

\title{Wavefront shaping of terahertz radiation using two-color flying-focus pulses with time-dependent focal velocities}

\author{A.L. Elliott}
\email{aellio18@ur.rochester.edu}
\affiliation{
University of Rochester, Laboratory for Laser Energetics, Rochester, New York 14623-1299 USA}

\author{H. Markland}
% \email{hmar@lle.rochester.edu}
\affiliation{
University of Rochester, Laboratory for Laser Energetics, Rochester, New York 14623-1299 USA}

\author{K.G. Miller}
% \email{kmill@lle.rochester.edu}
\affiliation{
University of Rochester, Laboratory for Laser Energetics, Rochester, New York 14623-1299 USA}

\author{S.W. Jolly}
\affiliation{
OPERA-Photonique, Université libre de Bruxelles, Brussels, Belgium}

\author{J.J. Pigeon}
% \email{jepi@lle.rochester.edu}
\affiliation{
University of Rochester, Laboratory for Laser Energetics, Rochester, New York 14623-1299 USA}

\author{J.P. Palastro}
\email{jpal@lle.rochester.edu}
\affiliation{
University of Rochester, Laboratory for Laser Energetics, Rochester, New York 14623-1299 USA}

\date{\today}

\begin{abstract}
Properly phased two-color laser pulses drive photoionization currents that emit broadband THz radiation. With a conventional two-color pulse, the ionization front travels at a nearly constant superluminal velocity, generating THz radiation with conical wavefronts at a Cherenkov-like angle. Here we show that the dynamic intensity peak of a two-color ultrashort flying-focus pulse can be used to control the shape of the THz wavefronts. Simulations demonstrate that non-uniform motion of the intensity peak and the ionization front it drives result in a time-dependent emission angle that determines the wavefront shape. A decelerating intensity peak, in particular, enables the generation of THz radiation with parabolic wavefronts that are well suited for collection and focusing.
\end{abstract}
     
\maketitle

\section{Introduction}

Terahertz (THz) radiation enables a wide range of applications, including spectroscopy, imaging, and ultrafast control of matter \cite{lewis_review_2014, hafez_intense_2016, fulop_laser-driven_2020,roadmap2023}. Many of these applications benefit from near-single-cycle THz pulses, such as those generated by the interaction of ultrashort laser pulses with solids, liquids, gases, or plasma \cite{Hebling2004,gopal_characterization_2013,jin_observation_2017,Liao2020,Bruhaug2024}. Gases and plasma, in particular, can withstand substantially higher laser-pulse intensities than solids or liquids, allowing for volumetric generation of high-field THz pulses \cite{hamster_subpicosecond_1993, yoshii_radiation_1997, loffler_gas-pressure_2002, Leemans2003, sheng_powerful_2005, damico_conical_2007, Antonsen2007, wu_single-cycle_2008,Miao2016,Pukhov2021, Fu2025, Kumar2025}. In the two-color approach, a laser pulse composed of an appropriately phased fundamental and second harmonic ionizes a gas, driving a photoionization current whose rapid evolution generates THz radiation \cite{kress_terahertz-pulse_2004,kim_terahertz_2007,kim_coherent_2008,Kim2009,babushkin_ultrafast_2010-1,You2012,berge_3d_2013,johnson_thz_2013,vaicaitis_influence_2018,nguyen_wavelength_2019,Stathopulos2024,simpson_spatiotemporal_2024,simpson_dephasingless_2024,Pigeon2026}. Typically, the radiation is emitted into a cone at a Cherenkov-like angle determined by the velocity of the ionization front \cite{You2012,johnson_thz_2013}. This approach is attractive because it can have a higher conversion efficiency than similar one-color schemes \cite{hamster_subpicosecond_1993,loffler_gas-pressure_2002,damico_conical_2007,Fu2025} and can take advantage of high-repetition-rate tabletop laser systems. Despite these advantages, conventional focusing of the biharmonic pulse constrains the ionization-front velocity and resulting THz emission angle, limiting tunability.

Flying focus techniques \cite{Sainte-Marie2017,froula_spatiotemporal_2018,jolly_controlling_2020,palastro_dephasingless_2020,caizergues_phase-locked_2020,Ramsey2023,ambat_programmable-trajectory_2023,pigeon_ultrabroadband_2024} produce laser pulses with an intensity peak that travels at a tunable velocity, enabling independent control of the ionization-front velocity and focal geometry \cite{palastro_ionization_2018,turnbull_ionization_2018,Howard2019,simpson_nonlinear_2020,Franke2021,Kabacinski2023,simpson_spatiotemporal_2024,simpson_dephasingless_2024,Fu2025,Pigeon2026}. These techniques create a moving focal point by tailoring the focal time and location of each frequency, temporal slice, or annulus of a pulse. In the ultrashort flying focus, an axiparabola focuses different annuli to different longitudinal locations \cite{smartsev_axiparabola_2019,geng_propagation_2022, oubrerie_axiparabola_2022}, while a radially stepped echelon sets the relative timing of the annuli \cite{palastro_dephasingless_2020,ambat_programmable-trajectory_2023,pigeon_ultrabroadband_2024}. Beyond enabling a tunable focal velocity, this configuration provides an extended focal range that far exceeds the Rayleigh range. These capabilities have been demonstrated in both simulations and experiments of two-color THz generation \cite{simpson_spatiotemporal_2024,simpson_dephasingless_2024,Pigeon2026}, where a constant-velocity flying focus was used to tune the ionization-front velocity and emission angle of the generated THz. 

\begin{figure*}
\includegraphics[width=0.9\textwidth]{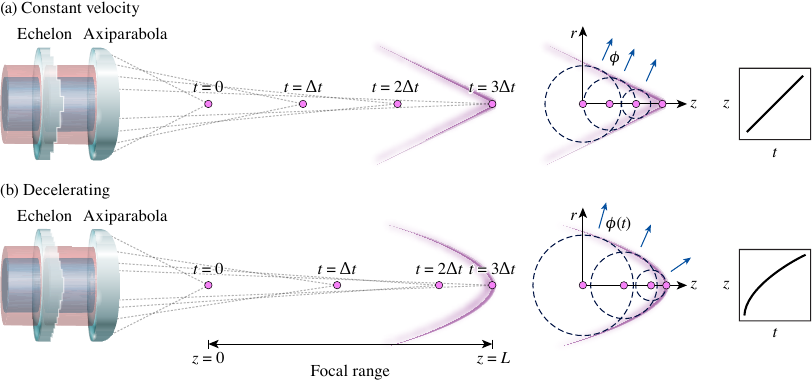}
\caption{Two-color THz generation driven by an ultrashort flying focus. A laser pulse composed of a fundamental and second harmonic is spatiotemporally structured by an echelon and axiparabola. The axiparabola produces an extended focal range, while the echelon is designed to produce a prescribed focal trajectory (rightmost plots). At the beginning of the focal range, the pulse ionizes a gas, initiating a moving ionization front. (a) A constant focal velocity yields a constant-velocity ionization front. The associated photocurrent emits THz radiation at a constant angle $\phi$, resulting in a conical wavefront. (b) A decelerating focal velocity yields a decelerating ionization front. The associated photocurrent emits THz radiation at a time-dependent angle $\phi(t)$, resulting in a parabolic wavefront. For illustrative purposes, the optical configuration is shown in transmission; in practice, the configuration would be reflective.}
\label{fig:f1}
\end{figure*}

The THz pulses created in two-color THz generation are typically single-cycle and extremely broadband, blurring the distinction between carrier and envelope. In this regime, the shape of the THz wavefront directly determines the spatiotemporal structure of the field. The wavefront, in turn, is set by the trajectory of the ionization front. For instance, a two-color flying-focus with a constant superluminal velocity produces THz radiation that is emitted off-axis with a conical wavefront, as illustrated in Fig. \ref{fig:f1}(a). This dependence of the wavefront on the ionization-front trajectory provides a means of shaping the spatiotemporal structure of the THz field.

Here, we demonstrate wavefront shaping of single-cycle THz radiation using two-color flying-focus pulses with non-uniform focal velocities. An analytic model is developed to determine the time-dependent ionization-front velocity required to generate a broad class of cylindrically symmetric wavefronts. For conical wavefronts, the model reproduces the constant-velocity Cherenkov condition. For parabolic wavefronts, the model yields an ionization front with a constant deceleration. The degree of deceleration determines the wavefront curvature and f-number of the generated THz. The model predictions are verified by simulations using the Unidirectional Pulse Propagation Equation (UPPE) \cite{kolesik_unidirectional_2002,kolesik_nonlinear_2004}. The model and simulations establish the ionization-front trajectory as a degree of freedom for structuring THz radiation, providing new opportunities to optimize the focused profile of THz radiation for applications.

\section{Wavefront Shaping}
\label{sec:model}

In two-color THz generation, an ultrashort laser pulse composed of a fundamental and second harmonic ionizes a gas. When the relative phase between the fundamental and second harmonic is close to $\pi/2$, the ionization produces a net current that generates THz radiation \cite{kim_terahertz_2007}. The THz is emitted at the Cherenkov angle $\phi=\arccos(v_\mathrm{T}/v_\mathrm{I})$ \cite{johnson_thz_2013}, where $v_\mathrm{I}$ is the velocity of the ionization front and $v_\mathrm{T}$ is the THz phase velocity. With conventional two-color THz configurations, the ionization front travels at a nearly constant superluminal velocity over limited ranges of the total propagation distance. The ionization-front velocity is determined by an interplay between the focal geometry, Kerr effect, ionization, and neutral dispersion. As a result, the emission properties cannot be tuned independently of the focal geometry. With ultrashort flying-focus configurations, the ionization front travels at the same velocity as the moving focus. This makes the ionization-front velocity an independent design parameter, providing greater control over the emitted THz.

Figure \ref{fig:f1} illustrates the concept of two-color THz generation driven by an ultrashort flying focus. A biharmonic laser pulse propagates along the positive $z$ axis and passes through an echelon and axiparabola. The echelon controls the relative delay $\tau(r)$ of each annulus, while the axiparabola controls its focal length $f(r)$. The focal-length profile is given by 
\begin{equation}\label{eq:axipar}
f(r) = f_0 + L - \bigg(\frac{r^2-r_0^2}{R^2-r_0^2}\bigg)L, 
\end{equation}
where $f_0$ is the nominal focal length, $r_0$ is the inner radius of the axiparabola, $R$ is the outer radius, and $L$ is the length of the focal range. For a given profile $f(r)$, the echelon is designed to impart the delay required to produce the time-dependent focal velocity $v_\mathrm{F}(t)$:
\begin{equation} \label{eq:taudot}
c\frac{d\tau}{dr} = \left[1 - \frac{v_\mathrm{F}(t_\mathrm{F}(r))}{c} + \frac{r^2}{2f^2(r)}\right]\frac{df}{dr},
\end{equation}
where
\begin{equation} \label{eq:tf}
t_\mathrm{F}(r) = \tau(r) + \frac{1}{c}\left[f(r) + \frac{r^2}{2f(r)} - 2s(r)\right]
\end{equation}
is the arrival time of each annulus at its focus and 
\begin{equation} \label{eq:sf}
s(r) = \frac{(R^2-r_0^2)}{4L}\ln\left[\frac{f_0+L}{f(r)}\right]
\end{equation}
is the sag function of the axiparabola \cite{ambat_programmable-trajectory_2023}.

After spatiotemporal structuring by the echelon and axiparabola, the biharmonic pulse propagates to the beginning of the focal range at $z=0$, where it ionizes a gas. The resulting ionization front travels at the focal velocity, $v_\mathrm{I}(t) = v_\mathrm{F}(t)$. When the focal velocity is constant (i.e., $dv_\mathrm{F}/dt=dv_\mathrm{I}/dt=0$), the THz radiation is emitted along the focal range at a constant angle $\phi =\arccos(v_\mathrm{T}/v_\mathrm{I})$, forming conical wavefronts [Fig.~\ref{fig:f1}(a)]. Thus, a constant focal velocity provides control over the emission angle independently of the focal geometry \cite{simpson_spatiotemporal_2024,simpson_dephasingless_2024,Pigeon2026}, but only produces conical wavefronts. A time-dependent focal velocity, in contrast, yields a time-dependent emission angle
\begin{equation}
\label{eq:angle}
    \phi(t)=\arccos\left[v_\mathrm{T}/v_\mathrm{I}(t)\right]
\end{equation}
that can be tailored to produce a prescribed wavefront shape [Fig.~\ref{fig:f1}(b)]. 

To determine the ionization-front trajectory $v_\mathrm{I}(t)$ required for a desired wavefront, the photoionization current is modeled as a line source on the $z$ axis, with each point on the line emitting at a specific time. Figure~\ref{fig:f1} depicts the line source as a sequence of point-like emitters (pink dots). Although each emitter radiates isotropically (dashed circles), the THz wavefront only forms where the emitted radiation constructively interferes (pink curves). 

Consider a THz wavefront in the near field, described by the cylindrically symmetric surface $z = z_\mathrm{w}(r)$ at $t = t_\mathrm{w}$, where $z_\mathrm{w} \gg L$. Each emitter contributes to the wavefront at a distinct radius $r$ and is identified by that $r$. The ray connecting an emitter on the $z$ axis at $z = z_\mathrm{e}(r)$ to the corresponding point on the surface is normal to the wavefront and makes an angle $\phi(t_\mathrm{e}(r))$ with the $z$ axis, where $t_\mathrm{e}(r)$ denotes the emission time. The emitter location can therefore be expressed in terms of the wavefront as
\begin{equation}
\label{eq:ze}
    z_\mathrm{e}(r) = z_\mathrm{w}(r) + r\left(\frac{dz_\mathrm{w}}{dr}\right)^{-1}. 
\end{equation}
The ray connecting an emitter to the wavefront reaches the wavefront after a time interval $t_\mathrm{w} - t_\mathrm{e}(r) = \ell(r)/v_\mathrm{T}$, where 
\begin{equation}
\label{eq:ray_length}  
    \ell(r) = \sqrt{r^2 + \left[ z_\mathrm{w}(r) - z_\mathrm{e}(r) \right]^2} 
\end{equation}
is the length of the ray. Combining Eqs.~\eqref{eq:ze} and ~\eqref{eq:ray_length} yields an expression for the emission time in terms of the wavefront:
\begin{equation}
\label{eq:te}  
    t_\mathrm{e}(r) = t_\mathrm{w} - \frac{r}{v_\mathrm{T}}\sqrt{1 + \left(\frac{dz_\mathrm{w}}{dr}\right)^{-2}}. 
\end{equation}
Given $z_\mathrm{w}(r)$, the required velocity is then
\begin{equation}
\label{eq:vI}  
    v_\mathrm{I}(t_\mathrm{e}(r)) = \frac{dz_\mathrm{e}}{dr}\left(\frac{dt_\mathrm{e}}{dr}\right)^{-1} = v_\mathrm{T}\sqrt{1 + \left(\frac{dz_\mathrm{w}}{dr}\right)^{2}}.
\end{equation}
Equation~\eqref{eq:vI} expresses the required ionization-front velocity parametrically in terms of $r$. To obtain $v_\mathrm{I}$ as a function of the emission time $t_\mathrm{e}$, Eq.~\eqref{eq:te} is inverted to find $r(t_\mathrm{e})$, which is then substituted into Eq.~\eqref{eq:vI}.

The preceding equations provide a straightforward procedure for determining the ionization-front trajectory that produces a prescribed wavefront. This procedure is now illustrated for three wavefront geometries. First, consider the case of a spherical wavefront described by the surface $z_\mathrm{w}(r) = (\mathcal{R}^2-r^2)^{1/2}$. Substituting $z_\mathrm{w}(r)$ into Eqs.~\eqref{eq:ze} and~\eqref{eq:te} yields $z_\mathrm{e} = 0$ and $t_\mathrm{e} = t_\mathrm{w} - \mathcal{R}/v_\mathrm{T}$, which is consistent with the expected result that point sources produce spherical wavefronts with radii determined by the retarded time, $\mathcal{R} = v_\mathrm{T}(t_\mathrm{w}-t_\mathrm{e})$. Next, consider a conical wavefront described by the surface $z_\mathrm{w}(r) = z_0 + \cot(\phi)r$, where $\phi$ is a constant. This is the geometry directly relevant to conventional two-color THz generation. In this case, Eq.~\eqref{eq:vI} recovers the Cherenkov relation $v_\mathrm{I} = v_\mathrm{T}/\cos(\phi)$ \cite{johnson_thz_2013}.

The remainder of the manuscript focuses on parabolic wavefronts, which are well suited to collection and focusing. The wavefront is described by the surface
\begin{equation}
\label{eq:para}
    z_\mathrm{w}(r)=\mathcal{Z}-\frac{r^2}{2\mathcal{R}}, 
\end{equation}
where $\mathcal{R}$ is the wavefront radius of curvature and $\mathcal{Z}$ is the longitudinal location of the vertex at time $t=t_\mathrm{w}$. Substituting Eq.~\eqref{eq:para} into Eqs.~\eqref{eq:te} and~\eqref{eq:vI} yields  
\begin{equation}
    \label{eq:vIpara}
    v_\mathrm{I}(t_\mathrm{e})=\frac{v_\mathrm{T}^2}{\mathcal{R}}(t_\mathrm{w}-t_\mathrm{e}).
\end{equation}
Equation~\eqref{eq:vIpara} reveals that parabolic wavefronts require an ionization front that undergoes a constant deceleration $-v_\mathrm{T}^2/\mathcal{R}$. The corresponding time-dependent emission location is then
\begin{equation}
    \label{eq:zepara}
    z_\mathrm{e}(t_\mathrm{e})= z_\mathrm{i} + \frac{v_\mathrm{T}^2}{2\mathcal{R}}[(t_\mathrm{w}-t_\mathrm{i})^2 - (t_\mathrm{w}-t_\mathrm{e})^2],
\end{equation}
where the integration constant is chosen so that the deceleration begins at $t_\mathrm{i}$, with the corresponding position $z_\mathrm{i} \equiv z_\mathrm{e}(t_\mathrm{i})$.

%The corresponding time-dependent emission location is then
%\begin{equation}
 %   \label{eq:zepara}
%    z_\mathrm{e}(t_\mathrm{e})=\frac{v_\mathrm{T}^2}{2\mathcal{R}}[t_\mathrm{w}^2 - (t_\mathrm{w}-t_\mathrm{e})^2],
%\end{equation}
%where the integration constant is chosen to place the initial emitter at the beginning of the focal range, i.e., $z_\mathrm{e}(t_\mathrm{e}=0) = 0$.

For echelon design and characterization of the THz, it is convenient to parameterize the ionization-front velocity in terms of $v_\mathrm{i} \equiv v_\mathrm{I}(t_\mathrm{i})$ and $v_\mathrm{f} \equiv v_\mathrm{I}(t_\mathrm{f})$, where $t_\mathrm{f}$ is the time at which the deceleration ends. With this parameterization, the minimum and maximum radii on the wavefront can be obtained from Eqs.~\eqref{eq:te} and~\eqref{eq:vI}:
\begin{align}
r_\mathrm{min} &= \mathcal{R}\sqrt{\left(\frac{v_\mathrm{f}}{v_\mathrm{T}}\right)^2-1} \\
r_\mathrm{max} &= \mathcal{R}\sqrt{\left(\frac{v_\mathrm{i}}{v_\mathrm{T}}\right)^2-1}. \label{eq:rmax}
\end{align}
These expressions reveal several important properties. First, the model presented here is limited to THz generation driven by superluminal ionization fronts, $v_\mathrm{I} \geq v_\mathrm{T}$. Second, a minimum radius $r_\mathrm{min} = 0$, corresponding to coherent on-axis emission, requires a final velocity $v_\mathrm{f} = v_\mathrm{T}$. Finally, the initial ionization-front velocity $v_\mathrm{i}$ controls the THz f-number $f_\#$. Identifying the radius of curvature as the distance from the virtual focus of the THz gives $f_\# = \mathcal{R}/(2r_\mathrm{max}$), or, equivalently
\begin{equation}\label{eq:fnum}
    f_\# = \frac{1}{2\sqrt{(v_\mathrm{i}/v_\mathrm{T})^2-1}}.
\end{equation}
Thus, for parabolic wavefronts, the divergence of the THz radiation is determined by the initial ionization-front velocity $v_\mathrm{i}$.

Equation~\eqref{eq:fnum} is sufficient to characterize the evolution of parabolic wavefronts during propagation. Equation~\eqref{eq:para}, however, specifies the wavefront at a particular vertex location $\mathcal{Z}$, with the corresponding radius of curvature $\mathcal{R}$. The location of the vertex follows from $\mathcal{Z} = z_\mathrm{i} + v_\mathrm{T}(t_\mathrm{w}-t_\mathrm{i})$ and Eq.~\eqref{eq:vI}, yielding $\mathcal{Z} = z_\mathrm{i} + v_\mathrm{i}\mathcal{R}/v_\mathrm{T}$. Similarly, the radius of curvature can be expressed in terms of the acceleration length $L_\mathrm{a} \equiv z_\mathrm{e}(t_\mathrm{f}) - z_\mathrm{i}$ using Eqs.~\eqref{eq:ze} and~\eqref{eq:vI}: \begin{equation}\label{eq:R}
    \mathcal{R}=\frac{2L_\mathrm{a}}{(v_\mathrm{i}/v_\mathrm{T})^2-1}.
\end{equation}
For the ionization-front velocities of interest, the initial velocity satisfies $v_\mathrm{i}\gtrsim v_\mathrm{T}$, so that $\mathcal{R} \gg L_\mathrm{a}$, confirming that the wavefront is specified in the near field, i.e., $\mathcal{Z} \approx \mathcal R\gg L_\mathrm{a}$. The wavefront at any other vertex location in the near field can be found using the linear relationship between the radius of curvature and vertex: $\mathcal{R}(\mathcal{Z}) = (v_\mathrm{T}/v_\mathrm{i})(\mathcal{Z}-z_\mathrm{i})$.

In practice, the formation of a THz pulse with a parabolic wavefront relies on uniform THz generation over the acceleration length, $L_\mathrm{a}$. Achieving uniform generation can be complicated by several effects: (1) longitudinal ramps or other inhomogeneities in the on-axis intensity profile produced by the axiparabola; (2) longitudinal variation of the spot size produced by the axiparabola \cite{Friberg96,oubrerie_axiparabola_2022}, which differs for the fundamental and second harmonic because it depends on the vacuum wavelength $\lambda$ (see Appendix \ref{app:axispot}); or (3) relative phase evolution between the fundamental and second harmonic due to neutral and plasma dispersion. To facilitate a uniform region of THz generation and promote the formation of a parabolic wavefront, the flying-focus trajectory is chosen so that the focal velocity is constant at $v_\mathrm{i}$ and $v_\mathrm{f}$ for adjustable intervals before and after the acceleration region, respectively. The corresponding piecewise focal velocity is then
\begin{equation}
\label{eq:trajectory_piecewise}
v_\mathrm{F}(t)=
\begin{cases}
    v_\mathrm{i} & 0 < t < t_\mathrm{i}  \\
    v_\mathrm{i} +  (\frac{t -t_\mathrm{i}}{t_\mathrm{f} -t_\mathrm{i}})(v_\mathrm{f} - v_\mathrm{i})
        & t_\mathrm{i} \le t < t_\mathrm{f} \\
    v_\mathrm{f} & t_\mathrm{f} \le t \le t_L,
\end{cases}
\end{equation}
where $t=0$ is defined when the focus arrives at the beginning of the focal region ($z=0$) and $t_L = t_\mathrm{f} + (L-L_\mathrm{a}-v_\mathrm{i}t_\mathrm{i})/v_\mathrm{f}$ is the time required for the focus to traverse the focal region. This velocity profile is used to design the echelon employed in the simulations described in the following section. 

\section{Simulations}
Wavefront shaping of THz radiation driven by two-color flying-focus pulses was demonstrated using the unidirectional pulse propagation equation (UPPE) \cite{kolesik_unidirectional_2002, kolesik_nonlinear_2004,couairon_practitioners_2011, berge_3d_2013}. The UPPE self-consistently evolves all resolved frequencies, making it well suited for modeling both the biharmonic driving pulse and the single-cycle, broadband THz radiation. As implemented here, the UPPE includes neutral dispersion to all orders, the third-order bound electron nonlinearity (i.e., $\chi^{(3)}$), the free-electron current density, and the depletion of electromagnetic energy due to ionization \cite{g_miller_spatiotemporal_2026}. See Appendix \ref{app:simulation} for details. 

The physical parameters used in the simulations are provided in Table \ref{tab:table1}. 
The parameters for the laser pulse were chosen to emulate commercially available multi-kHz Ti:sapphire laser systems. To improve the uniformity of THz generation, an axiparabola was used in which the outer radius focuses to the beginning of the focal region and the inner radius to the end [Eq.~\eqref{eq:axipar}]. This geometry results in a larger spot size and, consequently, more free electrons at the end of the focal region, enhancing THz generation where the focal velocity is lower. To mitigate longitudinal intensity modulations near the end of the focal region, the axiparabola and echelon were designed with a central hole of radius $r_0=1$ cm \cite{Friberg96}. The laser pulse incident on the optical assembly had a Gaussian temporal profile and a 20th-order super-Gaussian transverse profile with a Gaussian central hole matched to the central aperture of the axiparabola and echelon. In the far field, the pulse propagated through argon, with the frequency-dependent refractive index described by an empirically determined Sellmeier equation \cite{bideau-mehu_measurement_1981}. For each focal trajectory, the initial relative phase of the fundamental and second harmonic was optimized to maximize the THz energy. All decelerating trajectories included 0.75 cm regions of constant initial and final velocity, with $v_\mathrm{f} = c \approx v_\mathrm{T} =  0.99997\,c$ [Eq.~\eqref{eq:trajectory_piecewise}]. 

\begin{table}[t]
\caption{\label{tab:table1}
Simulation parameters for the drive laser pulse, optical configuration, and gas. The parameters for the fundamental and second harmonic are denoted by the subscripts 1 and 2, respectively.}
\begin{ruledtabular}
\begin{tabular}{lc}
Laser parameters & Value \\
\hline
$\lambda_1 ~(\mathrm{\upmu m})$ & 0.8 \\
$\lambda_1$ power $~(\mathrm{GW})$ & 16 \\
$\lambda_1$ duration $(\mathrm{fs})$ & 30 \\
$\lambda_2 ~(\mathrm{\upmu m})$ & 0.4 \\
$\lambda_2$ power $(\mathrm{GW})$ & 4 \\
$\lambda_2$ duration $(\mathrm{fs})$ & 30 \\
Total energy $\mathrm{(mJ)}$  & 0.6 \\
Super-Gaussian radial order & 20 \\
Super-Gaussian temporal order & 2 \\
\hline
Optics parameters & Value \\
\hline
Axiparabola $f_\#$ & 10 \\
Axiparabola radius $R$ (cm) & 5  \\
Hole radius $r_0$ (cm) & 1 \\
Focal range $L$ (cm) & 3 \\
Acceleration length $L_\mathrm{a}$ (cm) & 1.5 \\
\hline
Medium parameters & Value \\
\hline
Gas & Ar \\
Density $(\text{cm}^{-3})$ & $2.7 \times 10^{18}$  \\
Nonlinear refractive index $(\text{cm}^2 / \text{W})$ & $1 \times 10^{-20}$ \cite{zahedpour_measurement_2015} \\
Ionization energy (eV)  & 15.8 \\
Collision frequency (THz) & 10 \cite{sprangle_ultrashort_2004} \\
\end{tabular}
\end{ruledtabular}
\end{table}

\begin{figure}[t]
\includegraphics[width=1\linewidth]{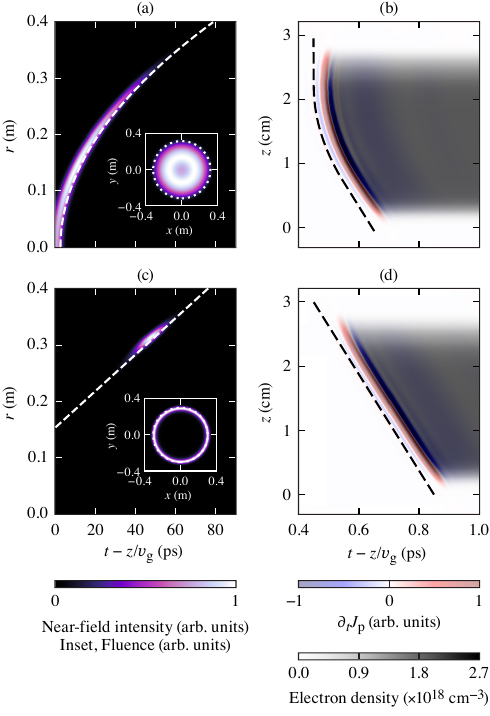}
\caption{Wavefront shaping of THz radiation using flying-focus--driven ionization fronts. (a) Near-field intensity profile of a THz pulse with a parabolic wavefront produced by a decelerating ionization front with $v_\mathrm{i}=1.004 \, c$ and $v_\mathrm{f}=c$, shown in (b). (c) Near-field intensity profile of a THz pulse with a conical wavefront produced by a constant-velocity ionization front with $v_\mathrm{F}=1.004 \, c$, shown in (d). In (a) and (c), the dashed lines indicate the prescribed wavefronts, $z_\mathrm{w}$.  The insets depict the transverse fluence profile. The dotted circles indicate the predicted maximum radius, given by Eq.~\eqref{eq:rmax} for the parabolic wavefront and by $r=\tan[\arccos(c/v_\mathrm{F})]z$ for the conical wavefront. Both near-field intensity profiles are shown at $z=3.5 ~\mathrm{m}$. In (b) and (d), the THz source, $\partial_t J_\mathrm{p}$ (blue-red scale), is overlaid on the electron density (gray scale), with the dashed lines tracing the prescribed flying-focus trajectory, $v_\mathrm{F}(t)$. The THz intensity and $\partial_t J_\mathrm{p}$ are filtered to retain frequencies between 5 and 20 THz.}
\label{fig:f2}
\end{figure}

Figures \ref{fig:f2}(a) and (b) illustrate THz wavefront control using a decelerating flying focus with an initial velocity $v_\mathrm{i}=1.004\,c$. As shown in panel (a), the near-field THz wavefront at $z = \mathcal{Z}$ has a parabolic profile in agreement with Eq.~\eqref{eq:para} (dashed line). The inset depicts the associated transverse fluence profile, with a dotted circle indicating the maximum radius $r_\mathrm{max}$ [Eq.~\eqref{eq:rmax}]. Panel (b) displays the decelerating ionization front that generates the parabolic wavefront. Both the ionization front (gray) and the differential plasma current density, $\partial_t J_\mathrm{p}$ (blue, red), which serves as the THz source, follow the prescribed flying-focus trajectory (dashed line). For comparison, Figs.~\ref{fig:f2}(c) and (d) show the same quantities for a constant-velocity flying focus with $v_\mathrm{F}=1.004\,c$. As expected, the constant-velocity flying focus produces a conical THz wavefront. The narrow ring-shaped fluence profile associated with the conical wavefront (inset) contrasts with the more uniform fluence profile of the decelerating trajectory. In both cases, $\partial_t J_\mathrm{p}$ and the THz wavefront are shown for frequencies between 5–20 THz. To improve visualization, the THz field has been convolved with a 2 ps Gaussian temporal profile.

\begin{figure}
\includegraphics[width=\linewidth]{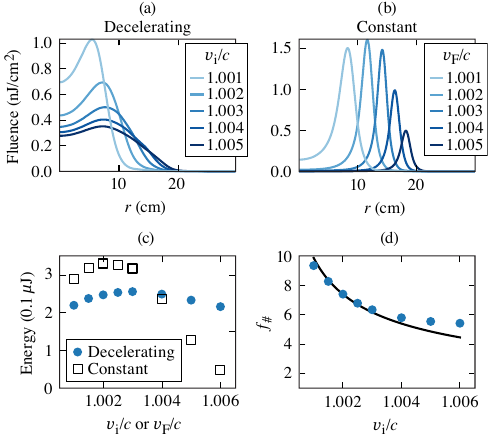}
\caption{Properties of THz radiation generated by uniformly decelerating and constant-velocity flying foci. (a) and (b) Near-field radial fluence profiles of THz generated by uniformly decelerating and constant-velocity flying foci, respectively. The decelerating trajectory produces a more uniform transverse fluence profile, consistent with a parabolic wavefront; the constant-velocity trajectory produces a more ring-shaped profile characteristic of a conical wavefront. (c) Total THz energy as a function of initial, $v_\mathrm{i}$, or constant velocity, $v_\mathrm{F}$. For the decelerating trajectory, the THz energy is relatively insensitive to the initial velocity. For the constant-velocity trajectory, the THz energy decreases at higher velocities because longitudinal chromatic aberration increases the pulse duration and reduces the peak intensity. (d) Simulated and analytically predicted THz f-number as a function of the initial velocity. The two are in good agreement up to $v_\mathrm{i} = 1.004  \, c$. Beyond this velocity, longitudinal chromatic aberration suppresses THz generation at the highest velocities, causing the simulated $f_\#$ to plateau. The fluence profiles and energies are shown for frequencies between 5 and 20 THz.}
\label{fig:f3}
\end{figure}

Figure \ref{fig:f3} compares the properties of THz radiation generated by decelerating flying-focus trajectories with different initial velocities to those generated by constant-velocity trajectories. In general, decelerating trajectories produce more uniform near-field fluence profiles than constant-velocity trajectories [cf. Figs.~\ref{fig:f3}(a) and (b)]. This occurs because the time-dependent emission angle distributes the THz radiation over detector radii ranging from $r= \tan(\phi_\mathrm{i})d$ to $r=0$, where $\phi_\mathrm{i} = \arccos(c/v_\mathrm{i})$ and $d$ is the longitudinal distance to the detector. The fixed emission angle of the constant-velocity trajectory, in contrast, concentrates the THz radiation into a single off-axis peak at $r=\tan(\phi_\mathrm{F})d$, where $\phi_\mathrm{F} = \arccos(c/v_\mathrm{F})$. In all cases, the fluence profiles are displayed in the near-field at $z = d = 1.78\,\mathrm{m}$. 

For constant-velocity trajectories, the energy emitted into THz radiation [Fig. \ref{fig:f3}(c)] follows the trend reported previously \cite{simpson_spatiotemporal_2024}. The emitted THz energy peaks when the focal velocity is close to the highest velocity of the axiparabola alone, $v_\mathrm{F} \approx 1+R^2/(2f_0^2) = 1.00125 \, c$, with the maximum energy slightly shifted to $v_\mathrm{F} \approx  1.002 \, c$ due to ionization refraction and collisional absorption. Away from the maximum, the energy drops because the radial group delay imparted by the echelon introduces longitudinal chromatic aberration. This increases the pulse duration and reduces the peak intensity, both of which decrease $\partial_tJ_\mathrm{p}$. By comparison, the THz energy generated by decelerating trajectories is relatively insensitive to $v_\mathrm{i}$. The decelerating trajectories sample a continuum of focal velocities, effectively averaging over the THz generation rates associated with those velocities.

Figure \ref{fig:f3}(d) highlights the tunable wavefront geometry enabled by a decelerating flying focus. Specifically, increasing the initial velocity of the flying-focus decreases the $f_\#$ of the emitted THz. The simulated $f_\#$ agrees with Eq.~\eqref{eq:fnum} up to $v_\mathrm{i} = 1.005 \, c$, beyond which it begins to plateau. The simulated f-number was calculated as $f_\# = d/(2w_\mathrm{T})$, where $w_\mathrm{T}$ is the Gaussian-equivalent spot size, defined as $\sqrt{2}$ times the root-mean-squared radius of the fluence.
% Suggested change to last sentence in paragraph: 
% The simulated f-number was calculated as $f_\# = d/(2w)$, where $w=\sqrt{2 \langle r^2 \rangle}$ is the Gaussian-equivalent spot size in terms of the root-mean-squared radius, $\sqrt{\langle r^2 \rangle}$.

The discrepancy between the theoretical and simulated $f_\#$ beyond $v_\mathrm{i} = 1.005 \, c$ arises from non-uniform THz generation along the acceleration length. The THz generation is sensitive to both the local ionization rate, which is set by the biharmonic field amplitude, and the relative phase between the fundamental and second harmonic at the ionization front. Variations in either quantity can locally enhance or suppress the THz generation, thereby accentuating or diminishing contributions from particular ionization-front velocities and emission angles. For constant-velocity trajectories, this effect is less pronounced because the THz is emitted at a fixed angle throughout the focal range. Although the flying-focus geometry was designed to mitigate nonuniform THz generation, variations arising from relative phase evolution remained and were more prominent for larger velocities. 

\begin{figure}
\includegraphics[width=1\linewidth]{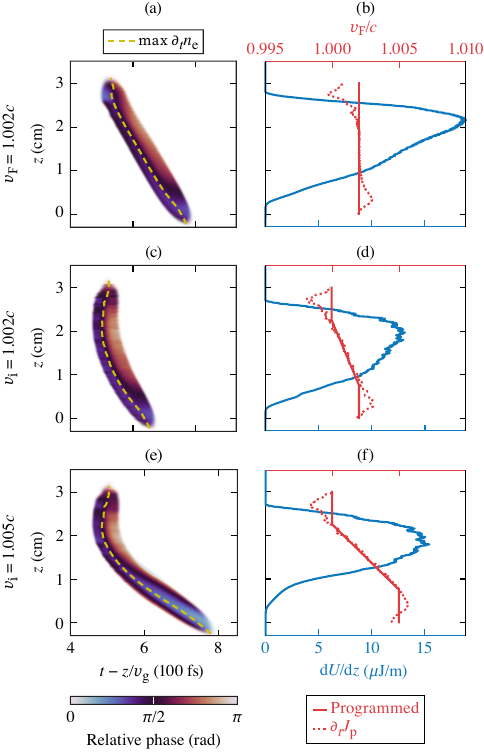}
\caption{Effect of the relative phase between the fundamental and second harmonic on the uniformity of THz generation. (left column) On-axis relative phase between the fundamental and second harmonic, masked where the fundamental intensity falls below 5\% of its maximum. The dashed yellow line indicates the ionization-front trajectory, defined by the location of the maximum ionization rate. (right column) Corresponding THz generation rate (blue, bottom axis), designed focal velocity (solid red, top axis) and simulated focal velocity (dashed red, top axis). The THz generation rate is highest when the relative phase is $\pi/2$ at the ionization front. In all three cases, the initial relative phase is chosen to maximize the total THz energy, with plasma dispersion evolving the relative phase to approximately $\pi/2$ along most of the ionization front. For $v_\mathrm{i} = 1.005 \, c$ (bottom row), the THz generation is suppressed early in the focal range where the velocity is high. The larger longitudinal chromatic aberration required at these velocities increases the pulse duration and decreases the intensity, leading to insufficient plasma generation to evolve the relative phase to $\pi/2$.}
\label{fig:f4}
\end{figure}

Figure \ref{fig:f4} examines how the relative phase evolution between the fundamental and second harmonic affects THz generation. The three rows correspond to the constant-velocity trajectory with $v_\mathrm{F} = 1.002\,c$ (top) and decelerating trajectories with initial velocities $v_\mathrm{i} = 1.002 \,c$ (middle) and $v_\mathrm{i} = 1.005\,c$ (bottom). The left column shows the trajectory of the ionization front (dashed line), defined by the location of the maximum ionization rate, overlaid on the on-axis ($r=0$) relative phase (color scale). The right column shows the rate of THz generation (bottom axis, blue), together with  the prescribed focal velocity (top axis, solid red) and the simulated ionization-front velocity (top axis, dotted red). The local relative phase is calculated as $\Delta \theta = \theta_2 - 2\theta_1 = {\arg[E_2 (E_1^*)^2]}$, where $E_j$ denotes the complex positive-frequency field of the $j$th harmonic and the asterisk denotes complex conjugation.

In all three panels, the THz generation is maximized when the ionization front coincides with the location where the relative phase equals $\pi/2$. Each bichromatic pulse is initialized with a relative phase that generally differs from $\pi/2$, so that ideally it evolves to $\pi/2$ at the ionization front, thereby maximizing the total THz energy. For the parameters considered here, this relative phase evolution is dominated by the plasma contribution to the refractive index. By comparison, the distance over which neutral dispersion advances the phase by $\pi$, $L_\mathrm{d}=(\lambda_1/4) |n(\lambda_1)-n(\lambda_2)|^{-1} = 27 \,\mathrm{cm}$, where $n$ is the neutral refractive index, is much longer than the focal range, $L = 3\,\mathrm{cm}$. 

%This is because the flying-focus pulse fully ionizes the relatively low density ($2.7\times10^{18} \, \mathrm{cm}^{-3}$) gas. 

For the constant-velocity trajectory with $v_\mathrm{F} = 1.002  \, c$ and the decelerating trajectory with $v_\mathrm{i} = 1.002  \, c$, plasma-induced phase evolution brings the relative phase to $\pi/2$ at the ionization front over most of the focal range [Figs~\ref{fig:f4}(a)--(d)]. This contrasts with the $v_\mathrm{i} = 1.005\,c$ flying focus, where early in the focal range, the intensity is too low to generate sufficient plasma for the relative phase to evolve to $\pi/2$ [Figs~\ref{fig:f4}(e) and (f)]. The reduced intensity results from the extended region of high velocity, where longitudinal chromatic aberration increases the duration and lowers the peak intensity, as discussed in the context of Fig.~\ref{fig:f3}(c). As a result, the THz generation is suppressed at the highest velocities, reducing the effective $r_\mathrm{max}$ [Eq.~\eqref{eq:rmax}] and leading to the saturation of the f-number observed in Fig.~\ref{fig:f3}(d).

%Effect of relative phase evolution between the fundamental and second harmonic on the uniformity of THz generation. harmonics, $\Delta \theta = \theta_2 - 2 \theta_1$, on THz generation over the focal range. Panels (a)-(b) correspond to a constant-velocity trajectory with $v_0=1.002\mathrm{c}$. Panels (c)-(d) show a piecewise decelerating trajectory described by Eq. (\ref{eqn:trajectory_piecewise}) with $v_\mathrm{i}=1.002\mathrm{c}$ and $\Delta v= -0.002\mathrm{c}$. Panels (e)-(f) show a trajectory with $v_\mathrm{i}=1.005\mathrm{c}$ and $\Delta v= -0.005\mathrm{c}$. Panels (a), (c), and (e) show $\Delta \theta$ as a function of propagation distance relative to the nominal focus, $z-f_0$, and moving-frame coordinate $t-z/v_\mathrm{g}$, where $v_\mathrm{g}$ is the first-harmonic group velocity. Transparency is smoothly applied to regions where the fundamental intensity is below approximately 5\% of its maximum value. The yellow dashed line indicates the location of peak $\partial_t n_\mathrm{e}$. Panels (b), (d), and (f) compare the programmed trajectory, the trajectory extracted from the maximum $\partial_t J_\mathrm{p}$, and the THz energy generation rate, $dU/dz$.

\section{Discussion \& Conclusions}

An ultrashort two-color flying-focus pulse can drive an ionization front with a time-dependent velocity, enabling the generation of THz radiation with a prescribed wavefront. The control over the near-field wavefront afforded by the time-dependent flying focus makes it possible to tailor the focused THz profile for specific applications. For instance, inverse-design techniques \cite{g_miller_spatiotemporal_2026} can determine the optimal focused THz profile, from which the required near-field wavefront and corresponding focal-velocity profile follow directly. This capability further extends the advantages of flying-focus pulses for two-color THz generation \cite{simpson_spatiotemporal_2024,simpson_dephasingless_2024,Pigeon2026}. In addition to wavefront shaping, the time-dependent flying focus retains the key features of constant-velocity flying-focus pulses: an ionization-front velocity and THz emission angle decoupled from the focal geometry, and an extended interaction length.  

An analytical model was developed to determine the time-dependent focal velocity required to generate a prescribed THz wavefront. Using the velocity profile from the model, an echelon can be designed to structure a two-color laser pulse with the radial group delay required to realize that profile. Parabolic THz wavefronts were considered as an example due to their utility in focusing and collection, with UPPE simulations confirming the analytical predictions. The generation of THz radiation with a parabolic wavefront requires a focal velocity that undergoes constant deceleration from a superluminal velocity to the THz phase velocity. The f-number of the resulting THz depends only on the initial superluminal velocity. 

More generally, the model can be used to generate THz radiation with arbitrary cylindrically symmetric wavefronts. For example, THz radiation with the wavefront produced by an axiparabola would have an extended focal range, with potential applications in THz electron acceleration or spatially resolved probing of matter. Future work could also explore extending the model to wavefronts produced by freeform optical elements.

%More generally, the model can be used to generate THz radiation with arbitrary cylindrically symmetric wavefronts. For example, THz radiation with an axiparabolic wavefront would have an extended focal range, providing extended spatiotemporal interactions for applications such as THz-driven electron-beam manipulation and spatially resolved probing of matter. Future work could also explore extending the model to wavefronts produced by freeform optical elements.

The practical realization of arbitrary wavefront shaping requires relatively uniform THz generation along the focal trajectory. The uniformity of THz generation can be improved by (1) using an axiparabola and echelon with a central hole to reduce on-axis intensity modulations, and (2) introducing intervals of constant velocity at the beginning and end of the focal trajectory so that the time-dependent portion occurs where the intensity is more uniform. Several design strategies could further improve the uniformity of THz generation. A ``phaser" optic, as described in Ref.~\cite{simpson_dephasingless_2024}, could tailor the nominal relative phase between the fundamental and second harmonic along the focal range to maintain a relative phase of $\pi/2$ at the ionization front. The transverse intensity profile of the pulse could be structured to compensate for the reduction in intensity caused by increased longitudinal chromatic aberration at higher velocities. Finally, the axiparabola could be replaced by a reflective axicon, which produces a constant spot size along its extended ``focal" range. Together with transverse structuring of the intensity profile, this could, in principle, create a near-uniform column of plasma with a nearly constant radius. Ultimately, the relatively low laser-pulse energies required for two-color THz generation allow for operation at high-repetition rates, where adaptive optics could be used to optimize both the total THz energy and the THz wavefront \cite{markland_rapidly_2026}. 

%Accurate wavefront shaping requires relatively uniform THz generation along the focal trajectory. 

\section*{Acknowledgements}
The authors would like to thank T.T. Simpson, D.H. Froula, and D. Ramsey for discussions. This material is based upon work supported by the Department of Energy [National Nuclear Security Administration] University of Rochester “National Inertial Confinement Fusion Program” under Award Number DE-NA0004144, U.S. Department of Energy, Office of Science, under Award Number DE-SC0021057, and the Air Force Office of Scientific Research under award number FA9550-24-1-0160.

\appendix
\section{Focal Spot Produced by an Axiparabola} \label{app:axispot}
An approximate expression for the transverse width of the central intensity core produced by the axiparabola can be obtained from the near-field f-number, $f(r)/(2r)$. Upon inverting $f(r)$ to express $r$ as a function of $f$, the transverse width is given by $w \approx \lambda f/[\pi r(f)]$. With the beginning of the focal range defined at $z=0$, the width becomes
\begin{equation}
w(z) \approx \frac{\lambda f_0}{\pi R}\left[1-\left(1-\frac{r_0^2}{R^2}\right)\frac{z}{L}\right]^{-1/2},
\end{equation}
where $f_0 \gg L$ has been used. This expression shows that the width monotonically increases along the focal range from $\lambda f_0/(\pi R)$ at $z=0$ to $\lambda f_0/(\pi r_0)$ at $z=L$.

\section{Simulation Details}
\label{app:simulation}

The simulations of two-color THz generation were performed using \textsc{super-jax} \cite{g_miller_spatiotemporal_2026}. The simulations proceed in two stages following the procedure outlined in Ref. \cite{simpson_dephasingless_2024}. In the first stage, a frequency-domain Fresnel integral propagates the two-color laser pulse from the optical assembly (i.e., the axiparabola and echelon) to the far field. In the second stage, the unidirectional pulse propagation equation (UPPE) \cite{kolesik_unidirectional_2002, kolesik_nonlinear_2004} propagates the pulse through the far field while simultaneously modeling THz generation and propagation. 

The first stage begins by initializing the transverse electric field of the two-color pulse immediately before the optical assembly as $E_0(\rho,\omega)$, where $\rho$ is the near-field radial coordinate. Immediately after the optical assembly ($z=-f_0$), the field is given by $E_0(\rho,\omega)e^{i\theta(\rho,\omega)}$, where  $\theta(\rho,\omega)$ is the phase imparted by the optical assembly. The field is then propagated through vacuum from $z=-f_0$ to the entrance of the gas at $z=z_\mathrm{g}$ using the frequency-domain Fresnel integral:
\begin{equation}
\label{eq:propagated_field}
\begin{aligned}
\tilde{E}_\mathrm{g}(r,\omega)
&= \frac{\omega}{icd_\mathrm{g}} e^{-i\omega d_\mathrm{g}/c}
\int \rho \mathrm{d}\rho E_0(\rho,\omega) \times \cdots \\
& \times J_0 \left( \frac{\omega r \rho}{c d_\mathrm{g}} \right)
\exp\left[
\frac{i\omega(r^2 + \rho^2)}{2cd_\mathrm{g}}
+ i \theta(\rho,\omega)
\right],
\end{aligned}
\end{equation}
\noindent where a tilde denotes a quantity in the frequency--radial-coordinate domain, $d_\mathrm{g} \equiv z_g + f_0$ is the distance to the entrance of the gas, $r$ is the far-field radial coordinate, and $J_0$ is the zeroth-order Bessel function of the first kind. The use of separate radial grids in the near and far fields when evaluating the Fresnel integral reduces the computational cost, particularly for smaller f-number geometries \cite{palastro_ionization_2018}.

The electric field obtained from the Fresnel integral, $\tilde{E}_\mathrm{g}(r,\omega)$, provides the initial condition for the second stage. In this stage, the UPPE evolves all resolved frequency components through the far field. These include the two-color laser pulse, the THz radiation, and any additional radiation generated through nonlinear frequency mixing. As implemented here, the UPPE is expressed as
\begin{equation}
    \dfrac{\partial \hat{E}}{\partial z} = i \left( k_z - \frac{\omega}{v_\mathrm{g}} \right) \hat{E} + \frac{\omega}{2 \varepsilon_0 c^2 k_z} (i \omega \hat{P}_\mathrm{g} - \hat{J}_\mathrm{p} - \hat{J}_\mathrm{i}),
\end{equation}
\noindent where $k_z=\sqrt{n^2(\omega)\omega^2/c^2-k_r^2}$ is the $z$-component of the wave vector, $n(\omega)$ is the linear refractive index, $k_r$ is the transverse wave vector, $v_\mathrm{g}$ is the group velocity of the fundamental frequency of the laser field, and a hat denotes a quantity in the frequency--transverse-wavevector domain. The shift of $k_z$ by $ \omega/v_\mathrm{g}$ transforms the time-domain coordinate from $t$ to the moving frame coordinate $\xi \equiv t - z/v_\mathrm{g}$. The terms on the right-hand side represent the nonlinear polarization density of the gas, $\hat{P}_\mathrm{g}$, the plasma current density, $\hat{J}_\mathrm{p}$, and an effective current density, $\hat{J}_\mathrm{i}$, that accounts for the electromagnetic energy lost during photoionization. These quantities are computed in the $\xi$ and spatial domains: 
\begin{align}
    P_\mathrm{g} &= \frac{4}{3}\varepsilon_0^2 c n_2 E^3, \label{eq:Pg} \\
    \partial_\xi J_\mathrm{p} &= -\nu_\mathrm{{en}} J_\mathrm{p} + \frac{e^2}{m_\mathrm{e}}n_\mathrm{e} E,
    \label{eqn:dJdt} \\
    J_\mathrm{i} &= \frac{w(E)n_\mathrm{n} U_\mathrm{I}}{E},
    \label{eq:Ji}
\end{align}
\noindent where $n_2$ is the nonlinear refractive index, $\nu_\mathrm{en}$ is the electron-neutral collision frequency, $n_\mathrm{e}$ is the electron density, $w(E)$ is the ionization rate, $n_\mathrm{n}$ is the neutral density, and $U_\mathrm{I}$ is the ionization energy. The electron and neutral densities evolve according to $\partial_\xi n_\mathrm{e} = w(E) n_\mathrm{n}$ and $n_\mathrm{n}=n_0-n_\mathrm{e}$, where $n_0$ is the initial neutral density. 

The UPPE system is solved in a cylindrically symmetric $\text{2D} + \xi$ geometry using quasidiscrete Hankel transforms \cite{guizar-sicairos_computation_2004} and a second-order predictor-corrector scheme for the nonlinear source terms \cite{g_miller_spatiotemporal_2026}. The Ammosov-Delone-Krainov (ADK) \cite{ammosov_tunnel_1986} ionization rate is used for $w(E)$, and a fixed electron-neutral collision frequency is employed in the current density equation to account for inverse bremsstrahlung absorption \cite{richardson2019}. The gas is modeled as singly ionizable, with the parameters used here given in Table \ref{tab:table1}. The refractive index is calculated using the Sellmeier equation from Ref. \cite{peck1964dispersion} and extrapolated to THz frequencies. All computations were performed using JAX \cite{jax2018}, NumPy \cite{harris2020numpy}, and SciPy \cite{virtanen2020scipy}. 

The temporal resolution and domain size were $\Delta \xi= 110~\mathrm{as}$ and $1.8~\mathrm{ps}$, corresponding to a spectral resolution of $\Delta f= 0.55~\mathrm{THz}$. The minimum radial resolution was $\Delta r= 1.7~\mathrm{\mu m}$, and the radial domain size was $4.5 ~\mathrm{mm}$. The longitudinal step size was $\Delta z= 21~\mathrm{\mu m}$. This resulted in a total number of grid points $(n_\xi,n_r,n_z) = (16384,2700,1834)$. The spatial and temporal resolutions were chosen to ensure numerical convergence, while the domain extents in $\xi$ and $r$ were chosen to mitigate aliasing and spurious reflections, respectively.

\bibliography{refs} 

\clearpage 

% \begin{figure}
% \includegraphics[width=1\linewidth]{Fig4_placeholder.JPG}
% \caption{Effect of a varying flying-focus spot size on THz generation over the focal range. (a) Time derivative of the plasma current density, $\partial_t J_p$ filtered to 5-20 THz as a function of radius and distance from the nominal focal point $z=f_0$. (c) Electron density as a function of radius and propagation distance. (b) Electron line charge as a function of propagation distance, calculated by integrating the traverse electron density from (c) over radius. (d) Two-color focal velocity as a function of propagation distance, showing a constant superluminal segment, followed by constant deceleration, and a constant luminal segment. This is compared with the rate of THz generation, $dU/dz$, by taking the THz electric field energy integrated over 5-20 THz, and taking a derivative with respect to $z-f_0$.}
% \label{fig:f5}
% \end{figure}

\end{document}